# Superconducting RF Cryomodule Demagnetization

Anthony C. Crawford     Fermilab Technical Div. / SRF Development Dept.     acc52@fnal.gov     05Mar15

This note presents measurements that support the proposition that it is feasible to de-magnetize a fully assembled superconducting RF cryomodule.

## The Cryomodule

A Fermilab ILC cryomodule has been partially assembled as shown in Figure 1.  The internal parts consist of one helium gas return pipe (HGRP), eight cavity helium vessels, eight single layer Cryoperm magnetic shields, and one Invar reference rod.  The steel vacuum vessel was previously de-magnetized as described in reference [1].  The magnetic shields and Invar rod had not been previously de-magnetized.

The individual cavity Cryoperm shields are open ended for this study.  It is therefore expected that there will be significant penetration of magnetic field into the shields for a distance of approximately one shield diameter ( ~23 cm).

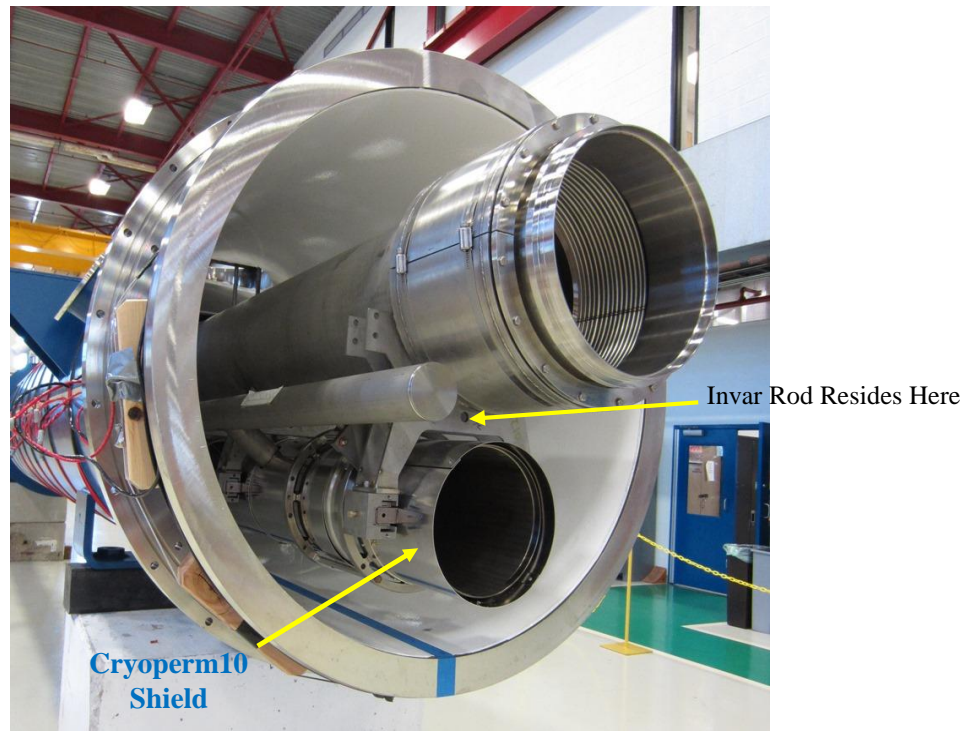

Figure 1.    Partially Assembled Cryomodule

The assembled cryomodule was de-magnetized (de-Gaussed) by exciting a series of coaxial coils with a decreasing bi-polar waveform. The coils are shown in Figure 2. The waveform is shown in Figure 3.

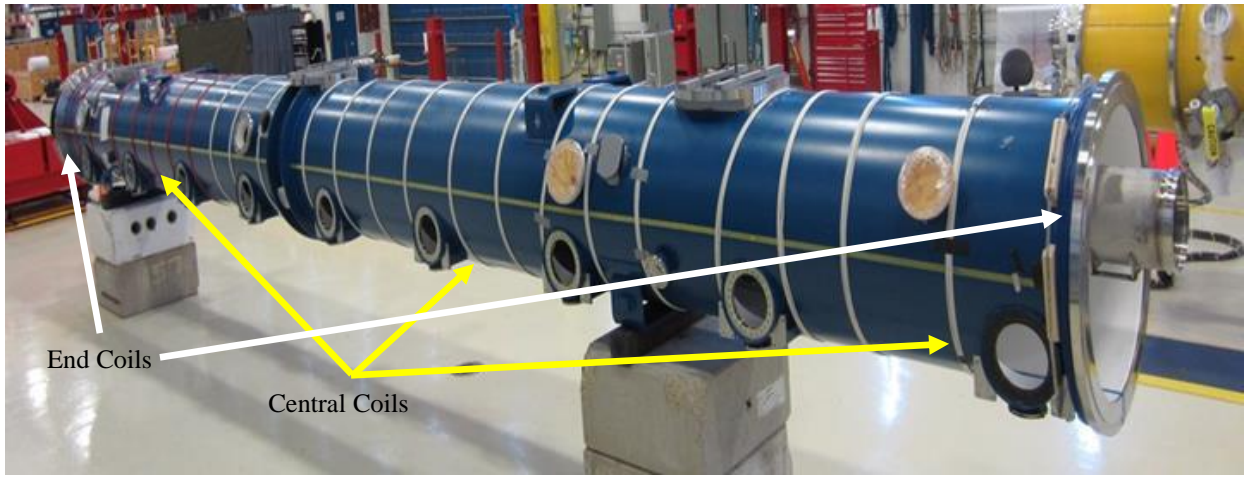

Figure 2.   Coils used for De-magnetization

The maximum desired current is entered into a LabView graphical user interface. Current as high as 65 Amperes is sourced from a Fermilab Booster Accelerator correction element power supply. The waveform progresses as follows:

| | |
|---|---|
| Linear ramp from zero current | 5 seconds |
| Initial current | 5 seconds flat top |
| Linear ramp to negative initial current | 5 seconds |
| Negative initial current | 5 seconds flat top |
| Linear ramp to 0.99 initial current | 5 seconds |

…   Et cetera

The amplitude of the flat top decrements by 1% of the initial current value for 100 cycles.

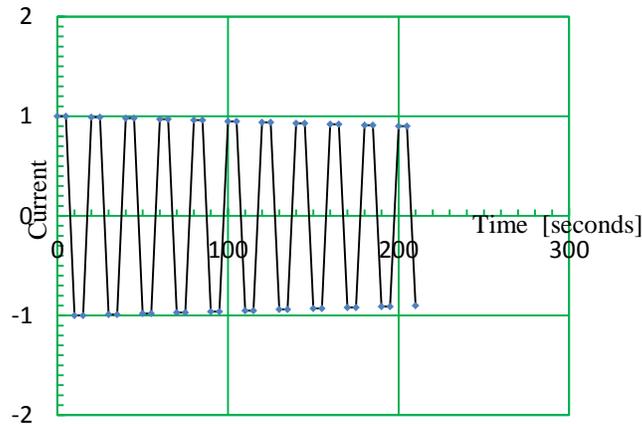

Figure 3.   Waveform Used for De-magnetization

The peak magnetomotive force density used for demagnetizing the central 90% of the cryomodule was 630 Ampere-turns/meter. Additional turns were used at the ends of the cryomodule in order to keep the magnetic field in the steel pipe from dropping by more than 25% relative to the value in the center of the cryomodule. The coil winding density for the central portion was 10 turns per meter.

**Results**

Results of the de-magnetization are shown in Figures 4 and 5. Field measurements were made along the axis of the cavity locations. The field plotted in the figures is the magnitude of the vector sum of three orthogonal field components. It should be noted that active field cancellation was not used during these measurements. The reason that the central fields are quite low is that the cryomodule was placed at a location with small (< ~50 mGauss) axial magnetic field, making the use of axial field cancellation unnecessary.

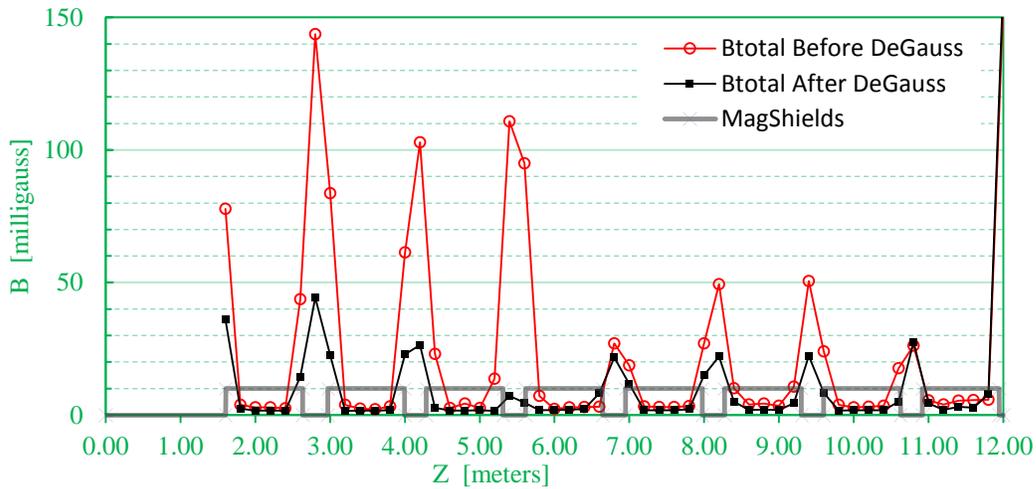

Figure 4.   Reduction of End Effects

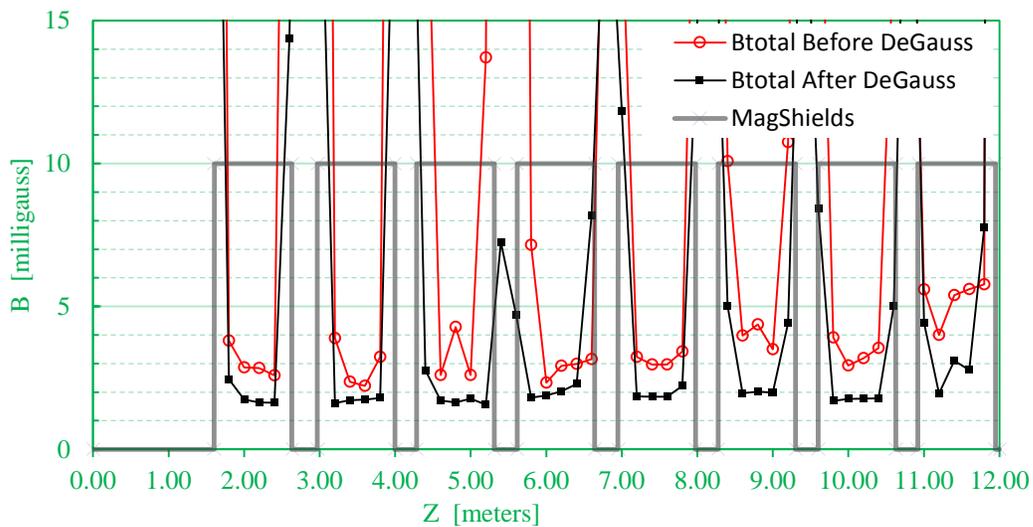

Figure 5.   Reduction of the Central Field

## Summary

1. Cryomodules can be de-magnetized in situ.
2. Both the Cryoperm shields and the Invar rod have significant remnant fields after handling and installation.
3. Both the Cryoperm shields and the Invar rod benefit from in situ de-magnetization.
4. This procedure is also relevant for cryomodules with stainless steel or aluminum vacuum vessels.
5. This procedure mitigates concern about re-magnetization during cryomodule handling and transport.
6. Cryomodules can be periodically de-magnetized in situ if it is suspected that re-magnetization has occurred.

## Future Studies

### De-magnetization Effects on Other Components

Before the de-magnetization procedure can be adopted for cryomodule production, it is necessary to verify that the presence of a quadrupole magnet and eight cavity tuner motors is acceptable. It is not anticipated that the quadrupole steel will be problematic. The stepper motors need to be checked for damage resulting in de-magnetization of their permanent magnets.

### De-magnetization as a Function of Temperature

The following is a quotation from reference [2]:

*"Cryoperm should be degaussed in situ with an amplitude slowly decreasing from a starting value of about 0.1 mT and a frequency of about 1 Hz to about 1 nT in 1 hr. This should be performed at the temperature and field environment in which the shield will be used, because temperature and field changes after degaussing the shield can substantially reduce its shielding properties."*

In order to investigate the temperature dependence of de-magnetization, we intend to measure the dynamic cryogenic load of the first pre-production LCLS2 cryomodule at the FNAL Cryomodule Test Facility both before and after cold, in situ de-magnetization of the cryomodule. However, the temperature of "cold" magnetic shields within a cryomodule typically decreases slowly with respect to the temperature of the cavities. Shield temperatures can be at values between 20K and 250K when the cavities transition to the superconducting state, depending on the thermodynamics and timing of the cryomodule cooldown. A reasonable compromise to an acceptable temperature for de-magnetization that covers this range may be necessary.